\documentstyle[aasms4,12pt]{article}
 
\begin{document}

\title{Discovery of a New $89$ Second X--ray Pulsar XTE J1906+09}
 
\author{D. Marsden, D. E. Gruber, W. A. Heindl, M. R. Pelling, 
\\ and R. E. Rothschild}
\affil{Center for Astrophysics and Space Sciences, University of 
California at San Diego \\ La Jolla, CA 92093} 

\begin{abstract}

We report on the discovery of a new pulsating X--ray source during 
{\it Rossi X--ray Timing Explorer} observations of a low galactic 
latitude field centered at RA (J2000) $= 19^{hr}05^{m}43^{s}$ and 
Dec (J2000)$=+08^{\circ}58\arcmin 48\arcsec$. Significant pulsations 
were detected by both the PCA and HEXTE instruments aboard {\it RXTE} 
at a fundamental period of $89.17\pm0.02$ seconds, with higher 
harmonics also visible in the $2-10$ keV power spectrum. The folded 
lightcurve from the source is multiply peaked at lower energies, and 
changes to single peaked morphology above $\sim 20$ keV. The phase 
averaged spectrum from the source is well fit by strongly absorbed power 
law or thermal bremsstrahlung spectral models of photon index $1.9\pm0.1$ 
or temperature $19.5\pm4.6$ keV, respectively. The mean neutral hydrogen 
column density is $N_{H}\approx10^{23}$ cm$^{-2}$, suggesting a distance 
of $>10$ kpc to the source and a minimum $2-10$ keV X--ray luminosity of 
$2\times10^{35}$ ergs s$^{-1}$. By comparison with other pulsars with 
similar periods and luminosities, we suggest that XTE J1906+09 has a 
supergiant companion with an underfilled Roche lobe. We speculate further 
that one of the M stars in a peculiar M star binary system may be the 
companion.

\end{abstract}
 

\section{Introduction}

The discovery of a new member of a small class of astrophysical sources 
is important because each source has the potential to constrain theoretical 
models of the source physics. Neutron star binary systems are an example 
of a small class of X--ray sources, with numbers presently totaling total 
$\sim200$. Accretion--powered pulsars can be divided into two broad 
classes based on their X--ray luminosities and spectra (\cite{white83}). 
The X--ray spectra of the higher luminosity sources 
($L_{X}\sim10^{36}-10^{37}$ ergs s$^{-1}$) are typically characterized by 
a hard power law of photon index $1<\Gamma<2$ out to an energy of $10-20$ 
keV, above which the spectrum ``breaks'' to a much steeper index 
($\Gamma>3$). The lower luminosity accreting X--ray pulsars tend not to 
have this broken power law shape, and instead have a softer spectral shape 
($\Gamma>2$) over a broad range of X--ray energies. Furthermore, 
observations of low luminosity burst sources have revealed weak power 
law components ($\Gamma\sim2$) when the inferred accretion accretion 
rate drops below a critical value (\cite{barret94}). In this {\it Letter} 
we present the detection of a new source which appears to be a low 
luminosity accreting X--ray source. 

\section{Data}

The $1^{\circ}$ (FWHM) field of view centered at the point RA (J2000) 
$=286^{\circ}.43$ and DEC (J2000) $=8^{\circ}.98$ was observed by 
the High Energy X--ray Timing Experiment (HEXTE) and the Proportional 
Counter Array (PCA) instruments aboard the {\it Rossi X--ray Timing 
Explorer} ({\it RXTE}) satellite on a number of occasions between 
August 16-19, 1996, with a total livetime of $\sim 25$ ksec. The target 
of the observations was the Interplanetary Network (IPN) error box from 
the soft gamma--ray repeater SGR 1900+14 (\cite{hurley94}), and the 
pointing was offset to avoid the bright confusing source 4U1907+097, 
which is $\sim0^{\circ}.7$ away. 

The HEXTE instrument consists of two clusters of collimated NaI/CsI 
phoswich detectors with a total net area of $\sim1600$ cm$^2$ and an 
effective energy range of $\sim15-250$ keV (\cite{roth98}). HEXTE 
background estimation utilizes near-real time observations of off-source 
viewing directions. The PCA instrument consists of $5$ collimated 
Xenon proportional counter detectors with a total net area of $7000$ 
cm$^{2}$ and an effective energy range of $2-60$ keV (\cite{jah96}). 
For the PCA instrument, the instrumental background estimate is 
determined from modeling blank sky observations to estimate both the 
internal background of the detectors and the background due to cosmic 
X-ray flux.

\section{Results}

\subsection{Timing Analysis}

Photon event times from the PCA in the energy range $2-10$ keV were 
extracted and binned into lightcurves of $1$ second binsize, after 
correcting the times to the solar system barycenter using the JPL 
DE200 ephemeris and {\it RXTE} pointing coordinates. The process was 
repeated for background data generated with the PCA background model, 
which was then subtracted from the on-source lightcurve. A 
Lomb-Scargle periodogram (\cite{press89}) was then formed from the 
data, resulting in the power spectrum shown in Figure 1 (top). Two 
harmonically spaced frequencies were detected at frequencies 
$f_{2}\sim 0.0224$ Hz and $f_{3}\sim 0.0336 $ Hz. The probability that 
each of these peaks is a random fluctuation is $\sim10^{-21}$, given 
the number of frequencies examined and the exponential probability 
distribution of the Lomb-Scargle periodogram. The presence of peaks  corresponding to $f_{5}$ and $f_{6}$ is also suggested by the 
periodogram, but at a marginal significance. Periodogram and FFT 
analysis using different time binsizes, no background subtraction, 
individual PCA detector units, and time interval subsets all show 
evidence for of one or two harmonic peaks in the power spectra. 

The harmonically-spaced frequencies $f_{2}$ and $f_{3}$ imply a 
fundamental frequency of $f_{1}\sim0.0112$, corresponding to a period 
of $\sim 89$ seconds. To search for the fundamental period we performed 
a chi-squared folding analysis on $128$ frequencies corresponding to 
periods about $89$ seconds, with a frequency resolution equal to the 
independent Fourier spacing implied by the total duration of the data. 
The result, shown in Figure 1 (bottom), indicates a fundamental period 
of $89.17\pm0.015$ seconds. We then searched the HEXTE event data for 
this periodicity, using the ${Z_{n}}^{2}$ test (\cite{buccheri83}). 
HEXTE on-source event times in the energy range $15-50$ keV were 
accumulated and corrected to the solar system barycenter using the the 
JPL DE200 ephemeris and {\it RXTE} pointing coordinates (as before). 
The $4\times10^{5}$ event times yielded a ${Z_{1}}^{2}$ value of $75.3$, 
using the single frequency corresponding to the period $89.17$ seconds. 
This corresponds to a random probability of $5\times10^{-17}$, given 
the probability distribution of ${Z_{n}}^{2}$. As a check, the $15-50$ 
keV HEXTE {\it off-source} events were analyzed in an identical manner 
and a value of ${Z_{1}}^{2}=0.3$ was obtained --- consistent with no 
periodicity.

The barycenter-corrected PCA and HEXTE event times were extracted for 
various energy ranges and folded on the fundamental period $P=89.17$ 
seconds (Figure 2). The corrected event times were normalized to 
MJD 50342.0 (TDB) before folding. The folded lightcurve (FLC) appears to 
change at energies between $20-30$ keV, with the pulse profile becoming 
more singly-peaked and possibly shifting in phase. In addition, the 
pulsed emission for energies above $\sim20$ keV peaks at a different 
phase than the emission at lower energies. Given the contributions from 
instrumental, cosmic X--ray, and Galactic ridge fluxes implied by the 
spectral fits (discussed below), we derive pulsed fractions of $16\%$, 
$6\%$, and $\sim 4\%$ in the energy ranges $2-10$, $10-20$ and $20-200$ 
keV, respectively.  

We have searched for evidence of Doppler shifts associated with possible 
binary motion in the observed periodicities from XTE J1906+09. Barycenter 
corrected $2-60$ keV PCA event times were collected for $7$ observation intervals throughout the observing period. The frequencies $f_{2}$ and 
$f_{3}$ were determined for each time interval via fitting of the 
peak in the ${Z_{1}}^{2}$ distribution, taken from small frequency ranges 
about $f_{2}=0.0224$ and $f_{3}=0.0336$, with a Gaussian. The fundamental
frequency $f_{1}$ (corresponding to the $89.17$ s period) was then obtained 
for each data set by the relation $f_{1}={1\over 5}(f_{2} + f_{3})$.
We find that the fundamental frequency obtained in this manner is 
consistent with a constant frequency over the $3$ day observation period. 
Because of the limited ($7$) number of observations and limited temporal 
coverage of the observations, however, the non-detection of significant 
Doppler shifts poses no serious constraint on the possible binary orbital  parameters of the system. For secular period changes, we fit a model to 
the data in which the period changes linearly with time, and obtained a 
$2\sigma$ upper limit of $|\dot{P}|< 2\times10^{-6}$.
 
\subsection{Spectral Analysis}

Spectral analysis of relatively faint X--ray sources at low Galactic 
latitudes is complicated by the presence of diffuse emission from the 
Galactic ridge. Recent work by Valinia \& Marshall (1998) indicates that 
the pointing direction of XTE J1906+09 is outside the bright, thin 
component of the ridge emission but inside the broader 
(FWHM$\sim4^{\circ}$) region. Only one of the HEXTE background positions 
is centered on a region of significantly higher Galactic X--ray flux 
$(l,b)=(39.6^{\circ},0.3^{\circ})$, and this background is excluded from 
the analysis. Following Valinia \& Marshall (1998), we model the Galactic 
ridge emission in the PCA with thermal and nonthermal components 
consisting of Raymond-Smith plasma (with solar elemental abundances and 
zero redshift) and power law spectral shapes, respectively, multiplied 
by a neutral hydrogen absorption factor. 

The phase-averaged PCA and HEXTE data were fit simultaneously to various 
continuum spectral models using XSPEC 10.0 and assuming that XTE J1906+09 
was at the center of the field of view. For the spectral fits, PCA and 
HEXTE data in the energy ranges $2.5-30$ and $15-250$ keV, respectively, 
were used. To account for uncertainties in the effective open area of the 
two instruments, the relative normalization of HEXTE was treated as a free 
parameter in the spectral fits. The plasma temperature, spectral index, 
and $N_{H}$ values of the Galactic ridge spectrum (PCA only) were held 
fixed at values appropriate for the ``Central Ridge'' (see Table 3 from 
Valinia \& Marshall 1998), but the normalizations were allowed to vary, 
to account for the fact that the {\it RXTE} pointing position is displaced 
from the Galactic plane by $1^{\circ}$. The results for a power law model 
for the XTE J1906+09 emission is given in Table 1. A thermal 
bremsstrahlung spectral shape of temperature $19.5\pm4.6$ keV and 
$N_{H}=(7.7\pm5.2)\times10^{22}$ cm$^{-2}$ fits the data equally well. 
More complicated spectral models (such as broken power laws) were not fit 
to the data, due to insufficient counts above $\sim 20$ keV. The reduced chi-squared values for both the power law and thermal bremsstrahlung 
spectral fits were $1.0$ for $523$ degrees of freedom. As expected, for 
both fits the Galactic ridge normalization values were lower than the 
values given in Valinia \& Marshall (1998) --- consistent with the 
$1^{\circ}$ offset from the Galactic plane --- and the relative 
normalization factor between the PCA and HEXTE was consistent with 
previous fits to high Galactic latitude sources (\cite{marsden97}). In 
the spectral fitting, we have assumed that all the observed iron line 
emission is from the Galactic ridge (Raymond-Smith plasma component), 
and therefore we do not attempt to constrain the iron line emission from 
XTE J1906+09.  

\section{Discussion}

Because of the low galactic latitude of the observed field, there is 
no shortage of possible XTE J1906+09 counterparts. The {\it ROSAT} 
all-sky survey (\cite{voges96}) sources RX J190724+0843.4 and RX 
J190717+0919.3 (\cite{vasisht94}) are unlikely counterparts, because 
their count rates are significantly higher (by at least an order of 
magnitude) than the predicted {\it ROSAT} PSPC count rate of XTE J1906+09, 
given the spectral parameters listed in Table 1. The predicted {\it ROSAT} 
HRI count rates are $7\times10^{-5}$ and $2.8\times10^{-4}$ counts s$^{-1}$ 
for the power law and thermal bremsstrahlung spectral forms, respectively. 
Supernova remnants such a G42.8+0.6 and other supernova remnants in the 
field of view (\cite{vasisht94}) are usually not associated with slow 
(period $>10$ s) pulsars because of age constraints. 

The pulsation period and inferred luminosity of XTE J1906+09 indicate 
that that the X--ray emission is powered by accretion onto a neutron star. 
The high absorption column density implied by the spectral fit suggests 
that the source is located at a distance of at least $\sim10$ kpc, and 
this yields a lower limit to the X--ray luminosity ($2-10$ keV) of 
$2\times10^{35}$ ergs s$^{-1}$. A variable pulse shape with energy 
(Fig. 2) is characteristic of accretion-powered neutron star systems, in 
which the emergent X--ray emission pattern is determined by the geometry 
and the distribution of matter in the vicinity. The scattering cross 
section is also energy dependent, and hence the pulse profile is expected 
to vary with energy. In addition, at lower energies more complicated 
pulsar lightcurves are expected due to photoelectric absorption by 
material around the neutron star. 

Assuming that XTE J1906+09 is an accretion-powered binary system, we 
can speculate on the nature of the system. The low luminosity and long 
spin period of XTE J1906+09 are atypical of low mass X--ray binary 
systems, but are consistent with a high mass system in which the 
accretion takes place via mass ejection from the companion star. Such 
high mass X--ray binaries can be divided into two classes: underfilled 
Roche lobe systems and Be-binary systems. Be-binary systems are 
transient systems in which the outflow from the Be star companion takes 
the form of either a stellar wind or an equatorial mass ejection 
(\cite{rapp82}). These sources are transient or highly variable, with 
X--ray luminosities ranging over $\sim4$ orders of magnitude  ($L_{X}\sim10^{34}-10^{38}$). If XTE J1906+09 is such a system, it 
probably has a long orbit ($P_{orb}>50$ days), due to the observed 
correlation between orbit and pulsar periods in Be-binaries 
(\cite{bildsten97}). One problem with this identification is that the 
X--ray spectra of these systems are characterized by power laws with 
photon indices $>3$ ($\sim20-100$ keV, \cite{bildsten97}). The 
$\sim20-250$ keV spectrum of XTE J1906+09 (using the HEXTE data only) 
is fit by a power law of photon index $1.45\pm0.45$, which is 
significantly harder than this. In addition, there are no known 
Be stars in the $1^{\circ}$ XTE J1906+09 error box (\cite{jas82}).

XTE J1906+09 could also be an underfilled Roche lobe supergiant system. 
These systems typically have long orbital periods ($P_{orb}>2$ days), 
long spin periods ($P_{spin}>100$ s), and relatively low X--ray 
luminosities of $10^{35}-10^{37}$ ergs s$^{-1}$ (\cite{bildsten97}). 
The inferred parameters of XTE J1906+09 are consistent with being a 
member of this class, but the X--ray spectral index of XTE J1906+09 
is flatter than that seen from Roche lobe supergiant systems 
(\cite{bildsten97}). 

The peculiar double M star infrared system discovered by Hartmann et al. 
(1995) is also in the {\it RXTE} field of view. The stars in this 
system are heavily reddened ($A_{V}\approx19.2$), implying a distance 
of $12-15$ kpc, and are thought to be gravitationally bound with an 
orbit of $2\times10^{6}$ years (\cite{vrba96}). Using the average 
extinction curve given in Savage \& Mathis (1979), and the relation 
between $N_{H}$ and $E(B-V)$ given in Korneef (1982), we derive the 
relation $A_{V}\approx 1.6N_{H22}$ mag, where $N_{H22}$ is $N_{H}$ in 
units of $10^{22}$ cm$^{-2}$. From the mean XTE J1906+09 $N_{H}$ value 
obtained from the power law and bremsstrahlung fits, we obtain 
$A_{V}=15\pm5$ mag, which is consistent with the value seen from these 
stars. 

Given the similarities between the observed properties of XTE J1906+09 
and high mass wind-fed X--ray binary systems, and the similar inferred 
distances of XTE J1906+09 and the M star supergiant system, we regard 
the latter as a promising counterpart candidate to XTE J1906+09. The 
large separation of the M stars would allow one to have a neutron star 
companion without significant effect on the M star system. Then the 
M star - neutron star system could be the source of the X--ray flux seen 
by {\it RXTE}. Vrba et al. (1996) have suggested that such a combination 
might be the counterpart of the soft gamma--ray repeater SGR 1900+14. 
The detection of $89$ second pulsations from these IR sources would verify 
the association of XTE J1906+09 with this system, and XTE J1906+09 would 
become the first known X-ray triple system. 

\section{Conclusions}  

We have presented the detection of a new $89$ second pulsating X--ray 
source with {\it RXTE}. The source was detected during observations of 
a region of Galactic plane containing the SGR 1900+14 error box. 
Characteristics of the folded lightcurves and spectrum suggest that 
the source, XTE J1906+09, is a low luminosity X--ray binary located 
beyond the Galactic center. The low luminosity and long pulse period 
of XTE J1906+09 indicate that this source is probably a high mass 
X--ray binary accreting via a stellar wind. We raise the possibility 
that the new source is associated with the highly absorbed double M 
supergiant binary system on the edge of the SGR 1900+14 error box. If 
this identification is correct, XTE J1906+09 is the first known X--ray 
triple system, and a candidate counterpart for SGR 1900+14.

\acknowledgments
 
We would like to thank the duty scientists at the online {\it RXTE} help 
desk for assistance in analyzing the PCA data. This work was funded by NASA grant NAS5-30720.

\clearpage

\begin{figure}
\caption{~The detection of periodicity. The top panel shows the Lomb 
periodogram of $2-10$ keV binned time series data from the PCA, with 
the harmonics labeled. The bottom panel shows the results of a 
chi-squared folding search for the fundamental period ($89$ seconds) 
of the pulsations, using the same data.}
\end{figure}

\begin{figure}
\caption{~The folded lightcurve of the $89$ second pulsations over 
a range of energies. The two middle panels correspond to the same 
energy range (see text). The background has not been subtracted.}
\end{figure}

\clearpage
 
\begin{deluxetable}{lcc}
\footnotesize
\tablewidth{0pt}
\tablecaption{Phase-Averaged Spectral Fit Results \label{Table 1}}
\tablehead{\colhead{Model Parameter} & \colhead{XTE J1906+09} & 
\colhead{Galactic Ridge}}
\startdata 
N$_{H}$\tablenotemark{a} & $10.8\pm4.0$ & $1.8$ (fixed) \nl
Photon Index & $1.9\pm0.1$ & $1.8$ (fixed) \nl
Power Law Flux\tablenotemark{b} & $16.0\pm2.5$ & $9.0\pm0.4$\nl
Plasma Temperature\tablenotemark{c} & $-$ & $2.9$ (fixed) \nl
Plasma Flux\tablenotemark{b} & $-$ & $7.7\pm0.7$ \nl
\enddata
\tablenotetext{a}{Neutral hydrogen absorption ($10^{22}$ H atoms cm$^{-2}$)}
\tablenotetext{b}{Unabsorbed $2-10$ keV flux ($10^{-12}$ ergs cm$^{-2}$ s$^{-1}$)}
\tablenotetext{c}{Raymond-Smith plasma temperature (keV)}
\end{deluxetable}


\begin{thebibliography}{}

\bibitem[Barret \& Vedrenne 1994]{barret94} Barret, D. \& Vedrenne, G. 
1994, \apjs, 92, 505 

\bibitem[Bildsten et al. 1997]{bildsten97} Bildsten, L. et al. 1997, 
\apjs, 113, 367

\bibitem[Buccheri et al. 1983]{buccheri83} Buccheri, R. et al. 1983, 
\aap, 128, 245

\bibitem[C\`{o}rdova 1995]{cordova95} C\`{o}rdova, F. A. 1995, 
in {\it X--ray Binaries}, ed. W. H. G. Lewin, J. van Paradijs, \& 
E. P. J. van den Heuvel (Cambridge: Cambridge University Press), 352

\bibitem[Hartmann et al. 1995]{hartmann95} Hartmann, D. H. et al. 1995, 
in AIP conf. Proc. 366, Workshop on High Velocity Neutron Stars, ed. R. 
E. Rothschild \& R. E. Lingenfelter (New York: AIP), 84

\bibitem[Hurley et al. 1994]{hurley94} Hurley, K. et al. 1994, \apjl, 
431, L31

\bibitem[Jahoda et al. 1996]{jah96} Jahoda, K. 
et al. 1996, EUV, X--ray, and Gamma--Ray Instrumentation for 
Astronomy VII, SPIE Proceedings, eds: O. H. V. Sigmund and M. Gumm, 
2808, 59

\bibitem[Jaschek \& Egret 1982]{jas82} Jaschek, M. and Egret, D. 1982, 
IAU Symposium No. 98, ``Be Stars'', eds. Jaschek, M. \& Groth, H.G., 
261

\bibitem[Koorneef 1982]{korneef82} Korneef, J. 1982, \aap, 107, 247

\bibitem[Marsden et al. 1997]{marsden97} Marsden, D. et al. 1997, 
\apjl, 491, L39

\bibitem[Press \& Rybicki 1989]{press89} Press, W. H. \& Rybicki, G. B. 
1989, \apj, 338, 277

\bibitem[Rappaport, S. \& Van den Heuvel 1982]{rapp82} Rappaport, S. \& 
Van den Heuvel, E.P.J. 1982, IAU Symposium No. 98, ``Be Stars'', eds. 
Jaschek, M. \& Groth, H.G., 327  

\bibitem[Rothschild et al. 1998]{roth98} Rothschild, R. E. et al. 1998, 
\apj, 496, 538

\bibitem[Savage \& Mathis 1979]{savage79} Savage, B. D. \& Mathis, J. S. 
1979, \araa, 17, 73

\bibitem[Valinia \& Marshall 1998]{valinia98} Valinia, A. \& Marshall, 
F. E. 1998, \apj, in press

\bibitem[Vasisht, Kulkarni, Frail, \& Greiner 1994]{vasisht94} 
Vasisht, G., Kulkarni, S. K., Frail, D. A. \& Greiner, J. 1994, 
\apjl, 431, L35

\bibitem[Voges et al. 1996]{voges96} Voges, W. et al. 1996, IAU Circular, 
6420

\bibitem[Vrba et al. 1996]{vrba96} Vrba, R. J. et al. 1996, \apj, 
468, 225
 
\bibitem[White, Swank, and Holt 1983]{white83} White, N. E., Swank, 
J. H., and Holt, S. S. 1983, \apj, 270, 711

\end{thebibliography}
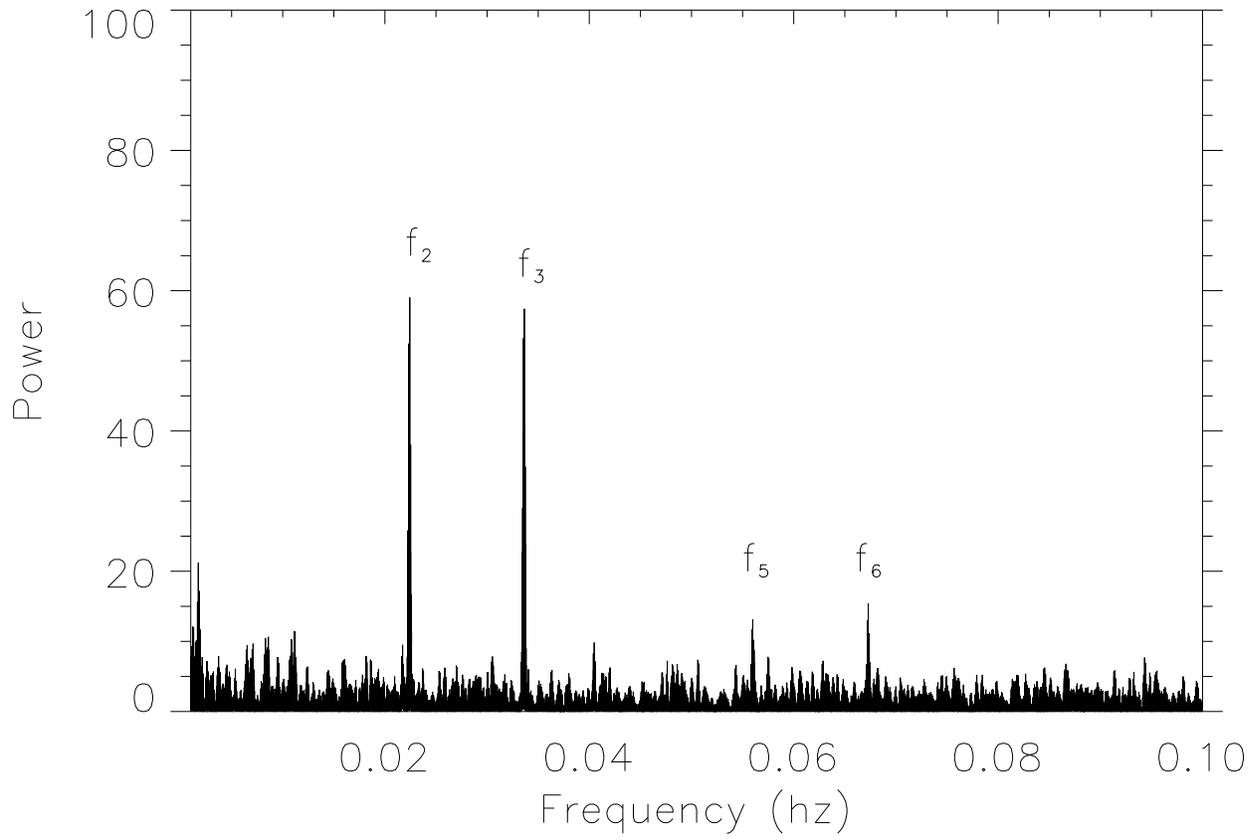
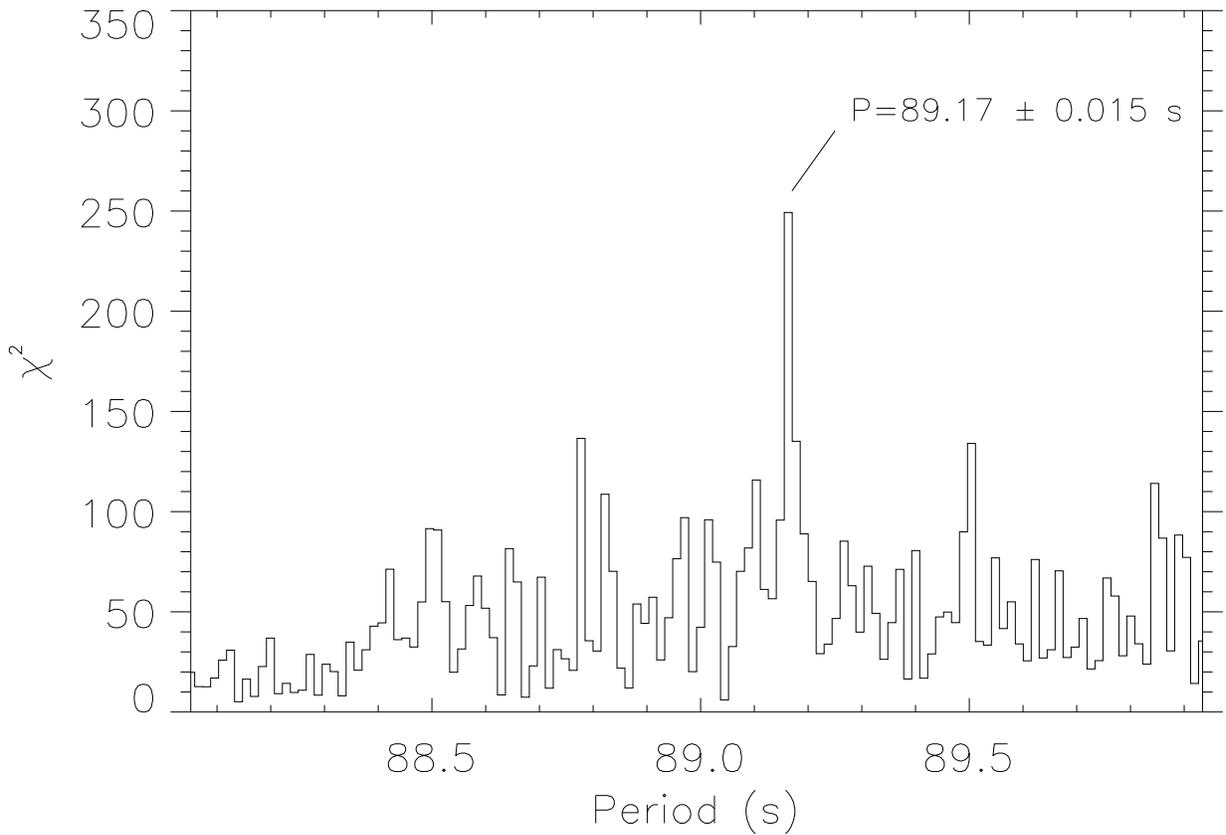
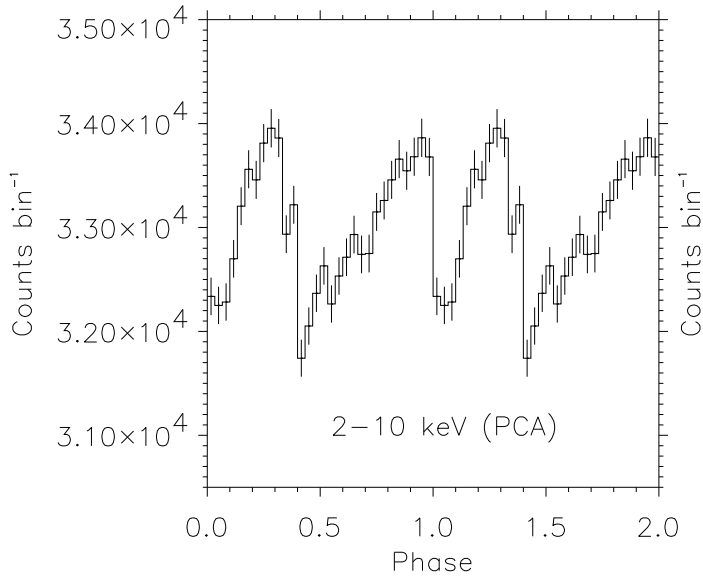
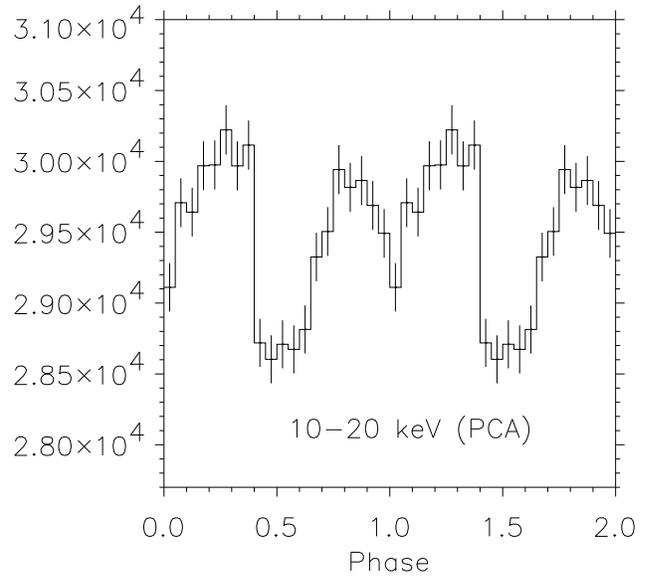
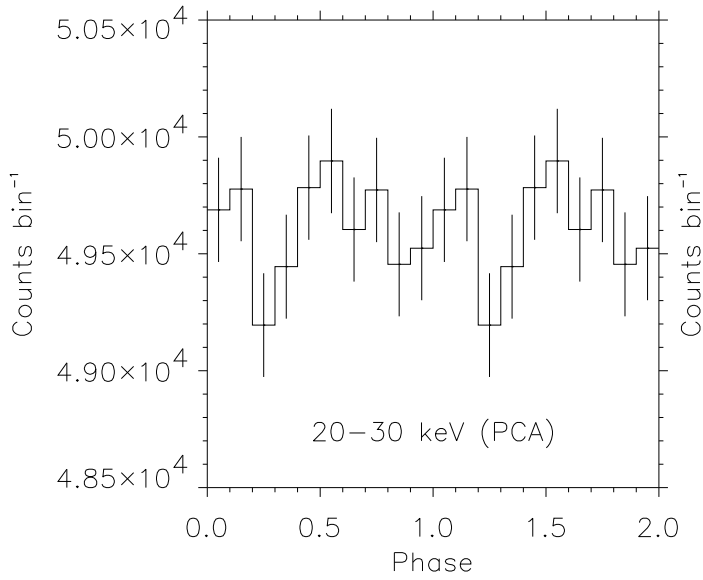
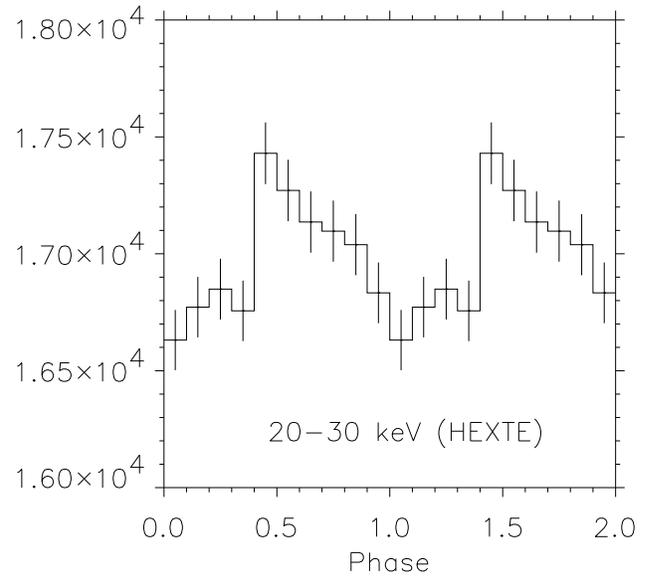
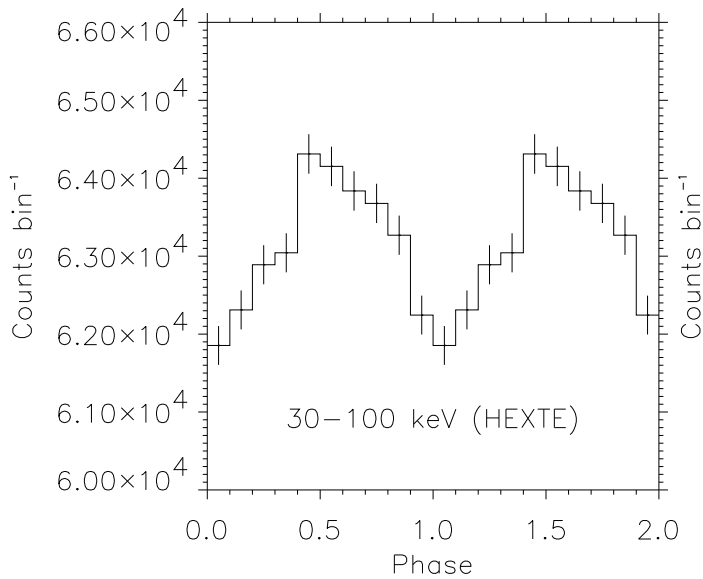
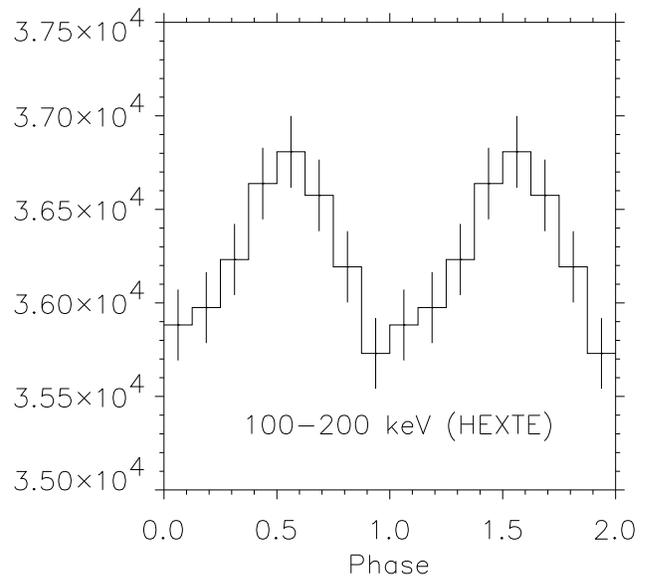
\end{document}